# Assessment of inter-model variability and biases of the global water cycle in CMIP3 climate models

**Beate G Liepert[1] and Michael Previdi[2]**

[1]NorthWest Research Associates, 4118 148th Ave NE, Redmond, WA 98052, USA

Email: Liepert@nwra.com

[2]Lamont-Doherty Earth Observatory of Columbia University, 61 Route 9W, Palisades NY 10964, USA

# Global water cycle assessment in CMIP3 climate models

**ABSTRACT**

Observed changes such as increasing global temperatures and the intensification of the global water cycle in the 20$^{th}$ century are also robust results of coupled general circulation models (CGCMs). In spite of this success model-to-model variability and biases that are small in first order climate responses however, have implications for climate predictability especially when multi-model means are used. We show that most climate simulations of 20$^{th}$ and 21$^{st}$ century A2 scenario performed with CMIP3 (Coupled Model Intercomparison Project Phase 3) models have deficiencies in simulating the global atmospheric moisture balance. Large biases of only a few models (some biases reach the simulated global precipitation changes in the 20$^{th}$ and 21$^{st}$ century) affect the multi-model mean global moisture budget and an imbalanced flux of -0.14 Sv exists whereas the multi-model median imbalance is only -0.02 Sv. For most models, the detected imbalances furthermore change over time. As a consequence, in 13 of the 18 CMIP3 models examined, global annual mean precipitation exceeds global evaporation, indicating that there should be a "leaking" (decrease) of moisture from the atmosphere whereas for the remaining 5 models a "flooding" is implied. Nonetheless, in all models, the actual atmospheric moisture content and its variability correctly increases during the course of the 20$^{th}$ and 21$^{st}$ centuries. These discrepancies therefore imply an unphysical and hence "ghost" sink / source of atmospheric moisture in the models whose atmospheres flood / leak. The source / sink of moisture can also be regarded as atmospheric latent heating / cooling and hence as positive / negative perturbations of the atmospheric energy budget or non-radiative forcings in the range of -1 to +6 W/m$^2$ (median is +0.1 W/m$^2$). The inter-model variability of the global atmospheric moisture transport from oceans to land areas, which impacts the terrestrial water cycle, is also quite high and ranges from 0.26 to 1.78 Sv. In the 21$^{st}$ century this transport to land increases by about 5% per century with a model-to-model range from 1 to 13% per century. We suggest that this variability is partially due to the different implementations of aerosol forcings in the models. The pole wards shifts of dry zones in climate simulations of the 21$^{st}$ century are also assessed in this study. It is shown that the multi-model means of the two subsets of models with negative and positive imbalances in the atmospheric moisture budget produce spatial shifts in the dry zone positions similar in size of the spatial shifts expected from 21$^{st}$ century global warming simulations. Thus, in this example of the dry zone extension, spatial multi-model means also depend on the selection of models which should be considered with caution in future analysis.

## 1. Introduction

As understanding of the climate system increased substantially over the years, modeling of the Earth's climate also progressed rapidly. Observed changes such as increasing global temperatures and the intensification of the water cycle are common, robust features of coupled general circulation models (CGCMs) as well as described in the 4$^{th}$ Assessment Report of the Intergovernmental Panel on Climate Change (IPCC-AR4). Due to these consistencies between models and observations, climate models are now widely used to predict climatic change under future anthropogenic emission scenarios. In addition to the robust results of global temperature and precipitation, CGCMs are more and more used to predict other complex climate responses of natural and anthropogenic perturbations. Such responses are for example shifts in the edge of the dry zones. Model-to-model variability that appears small in first order effects may however, have unexpected implications for more complex responses. An approach to overcome model uncertainties is suggested by Reichler and Kim (2008). They showed that for combinations of atmospheric variables, multi-model means of climate simulations represent the best estimates of the climate state when compared to 20$^{th}$ century observations. Their analysis was performed with data from the Coupled Model Intercomparison Project Phase 3 (CMIP3) archive. Calculating multi-model means of simulations and multi-model means of subsets of simulations became a common approach in recent years. Climate forcings, for example ozone recovery, and their climate responses were studied with multi-model means of hierarchies of models (e.g., Son et al., 2008). In this approach however, a few outlier models can skew outcomes significantly and can result in misleading conclusions. Another issue in predicting climate and source of uncertainty is that models are evaluated with 20$^{th}$ century observations and then used for predicting future climates. Tests that are designed independently of observations and observational uncertainty would hence be preferable.

Arguable the largest uncertainties in both climate observations and models stem from the hydrological cycle. While the basic processes are well known and the acceleration of the water cycle with global warming is well studied, the inter-model variability of the precipitation response and other hydrological processes remain high (Liepert and Previdi, 2009). Here we investigate these inter-model variability and biases of the global water cycle in CMIP3 models. We focus on the atmospheric branch of the hydrological cycle because it is intrinsically connected to the energy budget of the atmosphere and thus to climate forcings and feedbacks (see e.g. Liepert 2010). The atmospheric moisture content is by far the smallest storage term in the global water cycle. Although small variations in atmospheric moisture content can play key roles in the energy balance of the planet. Latent heating redistributes energy in the vertical column and cloud formation affects the emission of infrared and reflection of solar radiation while water vapor absorbs near-infrared and infrared radiation (e.g., Hansen *et al.* 1997, and Previdi and Liepert 2011). The atmospheric moisture transport from oceans to land constitutes the moisture input to the continental freshwater cycle. Hence the atmospheric, "oceans to land" moisture transport needs to be accurately predicted for reliable climate impact assessments. Another example of the importance of the atmospheric moisture balance is that the spatial distribution of the net amount of the fluxes of precipitation and evaporation identify the boundaries of the dry zones on Earth. These processes are investigated in this study. The manuscript first describes CMIP3 model data and data handling and in the second section introduces an analysis of biases of the global atmospheric moisture balance in climate models and its implications. In the second half of the study inter-model variability of atmospheric moisture transport from oceans to land and variability of extensions of dry zones are discussed.

**2. Climate Modeling Data**

The climate modeling experiments analyzed here are the archived CMIP3 simulations of fully coupled ocean-atmosphere general circulation models. Investigated in this study are the $20^{th}$ century scenarios with climate forcings determined by the individual modeling groups and the $21^{st}$ century scenario A2. Data from all runs of 18 models were downloaded from the archive (http://www-pcmdi.llnl.gov/ipcc/ipcc_data_status.php). The $21^{st}$ century runs were available for only 16 of the 18 models. The data sets of the $20^{th}$ and the $21^{st}$ century were combined into one data set. The models are listed in Table 1 and the abbreviations that follow the IPCC-AR4 nomenclature are also listed in the footnote of Table 1. For the atmospheric moisture balance analysis we examined ensemble means of all runs for each model and one arbitrary chosen run for each model. There were no differences in outcomes of the ensemble means versus the individual runs. Hence all results presented here are based on the analysis of individual model runs. For one model the analysis was performed with daily and as monthly data. The finer temporal resolution does not change the outcome of the assessment. Hence we use monthly datasets for this study. All datasets for each model are processed in the spatial resolution provided by the modeling groups and archived in CMIP3. No re-gridding was necessary for this analysis except for the calculation of the multi-model mean dry zone distributions. Available column integrated data were obtained from the archive and no vertical integration was performed, except water vapor for one model that did not provide column integrated water vapor for the archive (see Table 1). Individual land masks for each model were used for the calculation of the ocean to land atmospheric moisture transport. In case of models with mixed land-ocean grid cells, cells with land areas larger than 50% were considered land cells. The spatial integration was performed by summation of grid cell values, which were primary multiplied with the calculated grid cell areas. The variables we obtained are the monthly means of the surface latent heat flux and the monthly mean precipitation rates for each model. Precipitation consists of solid and liquid fluxes including negative values for dew and frost. Ocean evaporation was calculated from the available surface latent heat fluxes divided by the latent heat of vaporization ($L_{vap}$ = 2501 J/g) and sublimation of sea ice was calculated from the surface latent heat fluxes over sea ice divided by the latent heat of sublimation ($L_{sub}$ = 2835 J/g). Evapo-transpiration and sublimation over land was calculated from the surface latent heat fluxes divided by the latent heat of vaporization ($L_{vap}$ = 2501 J/g). This procedure slightly overestimates sublimation rates over land ice and snow because latent heat of vaporization is somewhat smaller than latent heat of sublimation. For one model we calculated land sublimation explicitly over ice and snow with the latent heat of sublimation and did not find accountable discrepancies after global integration. In the following we will use the expression "evaporation" for the sum of evapo-transpiration, evaporation and sublimation. From the CMIP3 archive we further obtained the archived monthly means of column-integrated water vapor for each model. For most models column integrated cloud liquid and ice water content was also available (marked in Table 1) and was added to the moisture content. The solid and liquid contributions to total atmospheric moisture are small compared to the contribution in the gas phase. Atmospheric moisture content was calculated as the sum of solid, liquid water, and water vapor.

**3. Global Atmospheric Moisture Balance**

    **3.1. Method**

According to Peixoto and Oort (1992) the moisture balance of an atmospheric column can be described in its vertically integrated form as follows:

$$\frac{\partial W}{\partial t} + \nabla_h \vec{Q} = E - P$$

$$\text{with} \quad \vec{Q} = \int_{z=0}^{\infty} \vec{v} \cdot q \, dz \quad \text{and} \quad W = \int_{z=0}^{\infty} q \, dz \tag{1}$$

117

118  The vector $\vec{v}$ is the horizontal wind velocity and $q$ the atmospheric moisture content (vapor, liquid, solid) of the

119  vertical layer $dz$. Integrated over the globe the horizontal moisture divergence $\nabla_h \vec{Q}$ in the atmospheric column

120  disappears and the column integrated atmospheric moisture storage change $\frac{\partial W}{\partial t}$ is balanced by the sources and

121  sinks of atmospheric moisture, which are the surface fluxes of evaporation minus precipitation $E - P$.

122  Furthermore, when applied to discrete data indexed $i$, the global atmospheric moisture gain or loss within the

123  time period of $n$ time steps can be described by the net accumulation of sources $E_i$ and sinks $P_i$ as indicated on

124  the right side of (2):

$$\frac{\partial W}{\partial t} = E - P \Rightarrow \langle W_n - W_1 \rangle = \sum_{i=1}^{n} \langle E_i - P_i \rangle \tag{2}$$

125

126  This means the global mean of the atmospheric moisture content changes with the net fluxes in and out of the

127  atmosphere. For monthly modeling data this means the annual atmospheric moisture gain/loss (e.g., from

128  January to December or any other 12-month period) is balanced by the yearlong net accumulation of monthly $E_i$

129  minus $P_i$.

$$\text{Res} = E - P - \frac{\partial W}{\partial t} \Rightarrow \text{Res}(y) \equiv \sum_{i=1}^{12} \langle E_i - P_i \rangle - \langle W_{12} - W_1 \rangle \tag{3}$$

130

131  Thus for the annual atmospheric moisture budget, a potential residual or imbalance $\text{Res}$ for the year $y$ can be

132  calculated and a time series of these annual residuals can be constructed.

133

134  **3.2. Results**

135  Figure 1 summarizes the 20[th] to 21[st] century long-term annual means and inter-annual variability of the two

136  components of the atmospheric moisture balance for each CMIP3 model. The inter-annual variability in all cases

137  is calculated as standard deviation of the annual means after the trends of the data records were removed. The

138  units for all moisture fluxes considered here are in Sverdrup (1 Sverdrup = 1 Sv = $10^6$ m$^3$ s$^{-1}$ = 31.6 *$10^{12}$ m$^3$ a$^{-1}$).

139 Illustrated in Figure 1 in red error bars is the inter-annual variability of the time series of $\frac{\partial W}{dt}$. Not recognizable

140 in Figure 1 are the mean annual atmospheric moisture changes. The long-term multi-model mean of moisture

141 change within the year is 0.00048 Sv. A positive, albeit small, increase like this is expected due to the increasing

142 moisture-holding capacity of the atmosphere with global warming. An increase in global atmospheric moisture

143 content $W$ of about 3.03 *$10^{12}$ m$^3$ over the 200-year period can be calculated from the multi-model mean of these

144 data. These changes in atmospheric moisture content are small compared to other storage terms of the water

145 cycle. Nonetheless the atmospheric moisture increases are important in the climate system because they intitate

146 the radiative water vapor and cloud feedback.

147 Also shown in Figure 1 are the long-term annual means of the time series of $E - P$ in columns and their inter-

148 annual variability in error bars, both in blue. Mean $E - P$ values exceed the actual variability of moisture storage

149 changes $\frac{\partial W}{dt}$ (red error bars) in almost all models and hence result in unbalanced moisture budgets (3). The

150 analysis was repeated with all available simulations and no differences in the results were obtained. As

151 mentioned before, the original data are only multiplied with grid cell areas before summation. Hence it is

152 unlikely that numerical errors from the integration can cause these residuals. Also mentioned in the data section

153 are missing cloud ice water content data for some models as well as the treatment of land sublimation as

154 evaporation. We further tested the possible biases due to these uncertainties with one model that includes all data

155 records. Omitting these data could not account for the observed deficiencies.

156 The long-term means, inter-annual variability and long-term trends of the residuals are also listed in Table 1. It is

157 clear that some models have balanced atmospheric moisture budgets. For example the model CGCM3.1 (T47),

158 which is flux corrected (marked with "*" in Table 1), closes the atmospheric moisture budget, while other

159 models (CSIRO-MK3.0 and UKMO-HadCM3) that are without flux adjustments are also in closure within the

160 uncertainty range (information on flux adjustments was taken from IPCC-AR4 2007 Table 8.1). Positive biases

161 could be identified for five and negative biases for thirteen models. Negative $E - P$ values in Table 1 and Figure

162 1 indicate a "leaking" ("drying") of moisture from the atmosphere and positive $E - P$ indicates a "flooding"

163 ("moistening") of the model atmospheres. The multi-model mean is negative with -0.14 Sv. The residuals are 164  generally small compared to the calculated global annual mean precipitation trends of the 20[th] and 21[st] century.

165  The results of each model are listed in Table 1. Although, the unphysical multi-model mean drying is about one

166  third the size of the multi-model mean precipitation trend (Table 1). This is the case because for a few models

167  the biases are large and can reach the magnitude of the inter-annual variability of precipitation. Consequently the

168  multi-model median of the residuals of -0.02 Sv is more appropriate. It is important to point out that the leaking

169  or flooding of the atmosphere that is anticipated based on the global imbalances of $E$ and $P$ is not reflected in the

170  actual simulated atmospheric moisture content $W$ and its inter-annual variation $\frac{\partial W}{\partial t}$ as discussed above and

171  shown in Figure 1. In the flooding models, the actually simulated moisture content changes $\frac{\partial W}{\partial t}$ are

172  significantly smaller than would be expected from the modeled $E - P$, whereas in the leaking models the

173  simulated increases of $\frac{\partial W}{\partial t}$ are of opposite sign than expected from the modeled $E - P$. This therefore implies

174  an artificial or unphysical source of atmospheric moisture in the models that leak and an unphysical sink in the

175  models that flood appear.

176  In general the global atmospheric moisture imbalances are small compared to precipitation and other fluxes in

177  the global water cycle. They are also not unexpected in climate models (see e.g., Rodriguez *et al.* 2010 and

178  Kavetski and Clark 2010). These small biases in atmospheric moisture however, become important if we

179  consider them as perturbation of the atmospheric energy budget. Additional moisture translates into excessive

180  latent heat release into the atmosphere through phase transition in precipitation formation (see also Edwards

181  2007). The atmosphere responds to this "ghost" latent heating with various feedback processes, which cannot be

182  identified easily. Hence the artificial source of moisture can be interpreted as "instantaneous", non-radiative

183  forcing of the energy balance. Table 1 lists the "excess latent heat" for each model, which are in the range of -1

184  to +6 W/m$^2$ with a small positive multi-model median of +0.1 W/m$^2$.

185  Climate predictions of water cycle strength however, are not necessarily influenced by biases, because

186  considered in climate model analyses are often differences of two climate states or changes over time from a

187  control run. Changes over time (drifts) of these biases however, would influence climate predictions of water

188  cycle strength. Tendencies of biases in each model calculated as linear trends for the annual residuals are listed

189 in Table 1. This analysis reveals drifts of the moisture imbalances. Tendencies, positive or negative, occur in the
190 residuals of most models as illustrated in Figure 2 where the initial imbalance of each run is removed. The drifts
191 are generally negligible compared to e.g., global precipitation trends (see Table 1). For some models however,
192 the trend in the residuals can be as large as five percent of predicted precipitation changes. For one climate
193 model the bias drift is double the precipitation trend of the 100-year period. Consequently multi-model means of
194 global precipitation trends should be considered with caution and multi-model medians are more reliable.

**4. Global Atmospheric Moisture Transport from Oceans to Land**

**4.1. Method**

As mentioned in the introduction the connection between land and ocean water cycles is important for many applications. The net atmospheric moisture transport from oceans to land connects ocean fresh water cycle and land hydrology. The water cycle is eventually closed by continental runoff, which is the return flow. Here we investigate the inter-model variability of the net atmospheric moisture transport from oceans to land to assess the uncertainty of this parameter of significant climatic change impact. Atmospheric moisture transport is commonly calculated as atmospheric moisture convergence with 3-dimensional wind and moisture fields (1). Because of the potentially high numerical uncertainty in calculating vertically integrated convergence directly we chose the more indirect atmospheric moisture budget approach. The atmospheric moisture supply for all land areas is derived from the moisture budget over the oceans.

$$\oiint_{\partial Oc} d\vec{S} \cdot \vec{Q} = \iiint_{Oc} \nabla \vec{Q}\, dV = \iiint_{Oc} \left( +E - P - \frac{\partial W}{\partial t} \right) dV \quad (4)$$

The left side of (4) describes the integration of the Gauss's flux theorem. Horizontal moisture flux $\vec{Q}$ through the surfaces $d\vec{S}$ of the entire atmospheric columns over the oceans is equal to the moisture convergence in the volumes $dV$ of the atmospheric columns over the oceans. This is the case because fluxes through the air-sea boundary are $E$ and $P$ while the fluxes at the top of the atmosphere are expected to be zero. The moisture convergence can then be replaced by the atmospheric moisture budget of (1). The globally integrated formulation of the atmospheric moisture transport from oceans to land is therefore:

214
$$\iint_{\partial Oc} d\vec{S} \cdot \vec{Q} = \left\langle \frac{\partial W}{\partial t} + E - P \right\rangle_{Oc} \quad (5)$$

215  The brackets $\langle \ \rangle_{Oc}$ symbolize the integration of all columns over the ocean areas. With (5), annual atmospheric

216  moisture transports from oceans to land can then be calculated for all CMIP3 climate models.

217

218  ### 4.2. Results

219  As pointed out in Table 1 the long-term, average atmospheric moisture transport from oceans to land varies quite

220  significantly from model to model with a range from 0.26Sv to 1.78Sv. The multi-model mean of 1.1 Sv and

221  median of 1.2 Sv for the $20^{th}$ and $21^{st}$ century simulations however, remain close to the observational estimate of

222  1.2Sv (e.g., Baumgartner and Reichel 1979). The inter-annual variability (calculated as standard deviation after

223  the record was detrended) is on average about 5-6% of the total transport in the models. In CMIP3 models, most

224  of the long-term variability stems from underlying trends towards increasing moisture transport to land areas.

225  The linear trends in moisture transports from oceans to land areas in the $20^{th}$ and $21^{st}$ century are shown in Figure

226  3. Overall, land areas will receive on average about 0.04-0.05 Sv (about 4%) more moisture per 100 years from

227  the oceans through atmospheric transport. This increase amounts to an extra moisture input to land of about the

228  size of the discharge of the River Nile per hundred years (e.g., Gupta 2007). In the models CCSM3 and ECHO-

229  G, the intensified moisture transports reach up to 0.13 Sv in 100 years or an increase of 10% of the total

230  atmospheric, ocean to land transport.

231  Further shown in Figure 3 are the linear trends of the residuals of the atmospheric moisture balances of

232  simulations of the $20^{th}$ and $21^{st}$ centuries (Table 1). In Figure 3 the two models (CNRM-CM3 and FGOALS-

233  g1.0) with reduced atmospheric moisture transport are also the models with strong increasing artificial leaking

234  from model atmospheres. The median of the drifts in atmospheric moisture balances is coincidently zero due to

235  compensation amongst models. Arguably drifts in the atmospheric moisture imbalance could affect the

236  atmospheric moisture transport from oceans to land areas because the largest flux in the global water cycle is

237  ocean evaporation. In general however, the trends in the global residuals are small compared to the trends in

238  atmospheric moisture transport (see multi-model mean and median in Figure 3).

239

## 5. Extension of the Dry Zone Edges

### 5.1. Method

Several approaches exist for identifying the edges of the dry zones in the subtropics (see Seidel *et al.* 2008 for an overview). For example, the descending branch of the Hadley circulation determines the edges of the subsidence region, which can be identified as the position of the jet streams or the zero net flow of mass from north to south in the lower atmosphere. Other distinct characteristics such as the stratospheric Brewer-Dobson circulation realized in the stratospheric ozone distribution or the tropopause height identify the width of the tropics and hence the edges of the dry zones as well. The ascending branch of the Hadley cell also produces cloud bands whose edges mark the beginning of the dry zone. The gradient of outgoing long-wave radiation is used for this approach. At the surface, dry zones are regions with evaporation exceeding precipitation and the edges are the zero contour line of $E - P$ or $E = P$. This definition has been used in observations and modeling studies (see e.g., Previdi and Liepert 2007).

### 5.2. Results

Figure 4a shows the $E = P$ contours for all CMIP3 models in light blue or light red color. Models with negative residuals in the moisture budget are marked in red and with positive residuals in blue. The inter-model variability of the dry zones is large. We further separate the models into two subsets. Dry zone edges in bold colors represent the multi-model composites of the two model subsets. The group with artificially leaking atmospheres is in bold red whereas the artificially flooding modeling group is marked in bold blue. As illustrated in Figure 4a the inter-model variability of the subsets is still significant. The "leaking model-composite" seems to reveal slightly narrower dry zone areas in the Northern and Southern Pacific particularly on the pole-ward edge compared to the flooding model-composite. The dry zones of the "leaking-models" are also slightly smaller in extent over the continents. Differences in the Eastern Pacific Walker circulation are recognizable with a zonal stretched descending branch in the tropical Pacific in the flooding models. The pattern indicates a more pronounced double ITCZ when the models' atmospheres experience positive biases.

In a former study (Previdi and Liepert 2007) we found as a robust result of all CMIP3 models the pole-ward extensions of the dry zone edges of about 1 degree of latitude on average with $21^{st}$ century global warming. The formerly published pole-ward extensions are reproduced for the CMIP3 models in Figure 4b as well. The mean spatial distributions of the dry zones of the first 20-year period and the last 20-year period of the 21$^{st}$ century are shown in Figure 4b in blue and green respectively with the multi-model mean contours in bold. The inter-model variability of the dry zone edges for the predictions is again quite large and the shifts in dry zone edges in the 21$^{st}$ century are similarly to the differences of the two subsets of the climate models with different biases. The differences and shifts can be clearer shown as zonal averages, because of the latitudinal structures of the $E - P$ fields. Zonal averaging was performed as follows: grid cell values with precipitation > evaporation ("wet" cells) were set to the discrete value 'zero' and grid cell values with evaporation $\geq$ precipitation ("dry" cells) were set to 'one'. These fields of discrete values were then zonal averaged. A zonal average of zero means the latitude band is outside the dry zone.

Figure 5 shows the results of this analysis for the combination of the negative and positive residual subsets and the two 20-year time intervals of the 21$^{st}$ century predictions of the models. The first and the last 20-year period of the leaking models are in solid and dashed lines, shown in red, and the first and last 20-year period of the flooding models are shown in solid (2001-2020) and dashed (2081-2100) lines shown in blue. The spreads between the models (not shown) are quite large for both, the biases and the prediction trends, particularly in the tropical belt. For the composites of the two modeling groups a narrower tropical rain belt (from around 15$^o$S to 15$^o$N) for the (red) leaking average s recognizable at the equator side of the southern hemisphere. The differences at the pole-ward edges of the subtropical dry zones are not as pronounced. Interestingly, for the (blue) flooding model composite the mid-latitude storm-track and polar region (50$^o$ – 80$^o$N and 60$^o$ – 80$^o$S) are more populated and hence drier compared to the leaking model atmospheres. In these zonal bands the model differences are larger then the 21$^{st}$ century shifts.

The pole-ward shift of the dry zone maximum extent is more prominent in the southern hemisphere. It is noteworthy that the overall magnitude of differences is similar in both comparisons but the climate change signal of shifting dry and wet zones, remains qualitatively the same for both bias groups. The strengths of the shifts are different between the subgroups.

We also calculated the areal extent of the dry zones as fraction of the total area of the globe, which indicates whether the dry zones are shrinking, expanding or just shifting. In Table 2 the areal fractions of the dry zones

Global water cycle assessment in CMIP3 climate models

294  show no significant differences between the two bias composites. The dry zones cover on average about 40% of
295  the globe in these models. The calculation is repeated with the A2 scenario global warming experiments. The
296  differences in areal sizes of the dry zones between the first two and the last two decades of the 21$^{st}$ century also
297  reveal no significant tendencies. In ten models the dry zones shrink slightly and in six models an equally small
298  extension (less than one percent) is indicated. The models with negative or positive net moisture budgets do not
299  show any preference for shrinking or expanding hence the areal fraction of the dry zones seems a robust feature
300  of the climate system similar to the global mean relative humidity.

301

302  **6. Discussion and Conclusions**

303  In this study we assessed global atmospheric moisture budgets in CMIP3 climate model simulations of the 20$^{th}$
304  and 21$^{st}$ century scenario A2. For these models Reichler and Kim (2008) showed that for combinations of
305  atmospheric variables the multi-model means represent the best estimates of the climate state when compared to
306  20$^{th}$ century observations. Based on our investigations, we suggest that for water cycle variables like
307  precipitation or $E - P$ however, a few models can bias multi-model means. The inter-model variability is reduced
308  when the multi-model median is used. We conclude this from the global atmospheric moisture budget, which is
309  out of balance by -0.14 Sverdrup in the multi-model mean whereas the multi-model median is only out of
310  balance by -0.02 Sv. The discrepancies in the moisture balance vary hugely amongst models and range from -
311  1.34 to 0.20 Sv. The biases are also not constant over time and can drift significantly. Positive and negative drifts
312  were detected for the simulations of the 20$^{th}$ and 21$^{st}$ century. The trends in the model biases range from a few
313  percent of simulated global precipitation trends to less than a tenth of a percent. For one model the trend in the
314  bias is 200% of the predicted precipitation change in a 100-year period. Hence models with large drifts should be
315  excluded from multi-model mean calculations.
316  The global biases in moisture balance can also be regarded as artificial "leaking" of moisture from the
317  atmosphere when the imbalances are negative (which is the case in thirteen of the eighteen models).  This
318  leaking is artificial in the sense that the actual moisture content of the atmosphere is simulated to increase during
319  the 20$^{th}$ and 21$^{st}$ centuries. Thus, discrepancies between the simulated $\frac{\partial W}{\partial t}$ and $E - P$ implies an unphysical, 320  "ghost" source of moisture. This "ghost" moisture source can also be described as excess latent heating in the
321  energy budget of the atmosphere and therefore as climate perturbation or non-radiative "ghost" forcing. The
322  excess latent heating ranges from -1 to +6 W/m$^2$, with a small multi-model median of +0.1 W/m$^2$. Radiative
323  forcings of non-$CO_2$ greenhouse gases are in the same size range.
324  The model-to-model variability of atmospheric moisture transport from oceans to land is quite high and ranges
325  from 0.26 to 1.78 Sv, while the inter-annual standard deviations are around 0.06 Sv in the 20$^{th}$ and 21$^{st}$ century
326  scenarios. The global flux is 1.2 ± 0.3 Sverdrup (Sv) in the multi-model mean, which matches observations quite
327  well (Baumgartner and Reichel, 1975; Schanze et al., 2010). The changes over the 100-year time period are all
328  positive (except the two models with strongly drifting imbalances). Based on our study we expect that land areas
329  receive on average about 4% more moisture in the next century with 0.08 Sv in the 21$^{st}$ century. This is about
330  half of the observed drainage of a major river such as the Lena in Siberia that has an annual runoff of 0.17 Sv
331  (see e.g., Gupta 2007). A reason for the inter-model variability of the moisture transport might be representation
332  of non-GHG forcings in the models. In Liepert and Previdi (2009) we suggested that some of the discrepancies
333  noted in precipitation changes between models and observations may be due to the various ways natural and
334  anthropogenic aerosols are treated in IPCC-AR4 climate models. Here we use the same hypothesis for the
335  moisture transport. Two special 20$^{th}$ century runs of the GISS-ER fully coupled model (prepared for IPCC-AR4)
336  were used where anthropogenic forcings were introduced individually (see Hansen *et al.* 2005). In this model
337  version the global atmospheric moisture budget is almost balanced and no drift over time could be detected. The
338  atmospheric moisture transports from oceans to land is calculated from outputs of the control (i.e., unforced) run
339  'CTL', the 20$^{th}$ century anthropogenic greenhouse gases only 'GHG' and of the 20$^{th}$ century anthropogenic
340  aerosols 'AER' (including direct and indirect effects) experiments. The aerosol forcing is spatially and
341  temporally non-homogeneous. The GISS model results lie well within the variability of the models shown in
342  Figure 3. The resulting atmospheric moisture transports from oceans to land in the 20$^{th}$ century are 0.90 Sv in the
343  control run, 0.95 Sv in the GHG and 0.99 Sv in the Aerosol experiment. The moisture transport to land areas
344  increases with 0.13 Sv per 100 years in the GHG and only 0.06 Sv per 100 years in the Aerosol experiment in
345  spite of a similar sized, albeit opposite in sign, response of the surface radiative energy budget (Romanou et al.,
346  2007). This result suggests that the treatment of aerosols in climate models affects the atmospheric moisture transport from oceans to land differently than GHG forcings and may cause some of the model-to-model variability.

The analysis of the dry zone extension reveals large model-to-model variability as well. The differences in spatial pattern between model groups with positive and negative moisture imbalances are comparable to the differences in the predicted pattern due to climate changes in the 21$^{st}$ century. Simulated pole-ward shifts of dry zones hence are dependent on the selection of models used for the analysis and these multi-model assessments should be evaluated with caution. The current study has not addressed the possible causes of the biases, which is beyond the scope of the manuscript. This question is subject of ongoing and future research and will be published at a later time.

*Acknowledgements:*

The authors thank the modeling groups for making their model output available for analysis, the Program for Climate Model Diagnosis and Intercomparison (PCMDI) for collecting and archiving this data, and the WCRP's Working Group on Coupled Modeling (WGCM) for organizing the model data analysis activity. The WCRP CMIP3 multi-model dataset is supported by the Office of Science, U.S. Department of Energy. This work was sponsored by NASA-MAP Program grant #NNG06GC66G and NSF Antarctic Oceans and Atmospheric Sciences grant #0944103. We thank Reto Ruedi, and Ken Lo for providing the NASA-GISS model simulations and the reviewers for their insightful comments.

**Table 1.** Global annual means, inter-annual variability and trends of residuals of the atmospheric moisture balance $Res = \left(E - P - \frac{\partial W}{\partial t}\right)$ as described in (3) for CMIP3 climate models. Listed are global annual means and standard deviations for the model time series of the 20$^{th}$ and 21$^{st}$ scenario A2. Also listed are global precipitation *(P)* trends and the percentage of residual *(Res)* trend to precipitation *(P)* trend for the same simulations. The excess latent heating that corresponds to the moisture imbalance is given in Wm$^{-2}$. Models with flux correction are marked with *, no cloud ice data available **, and models without cloud ice/*water* data are marked with ***.

| CMIP3 Model* | Mean ± Std.dev. $Res = \left(E - P - \frac{\partial W}{\partial t}\right)$ (Sv) | Trend $Res = \left(E - P - \frac{\partial W}{\partial t}\right)$ (Sv/100yr) | Trend P (Sv/100yr) | Trend Res / Trend P % | Excess Latent Heat (Wm$^{-2}$) |
|---|---|---|---|---|---|
| BCCR-BCM2.0 | -0.453 ± 0.010 | -0.010 | 0.41 | -2.44 | -2.22 |
| CCSM3 | -0.018 ± 0.006 | -0.000 | 0.53 | -0.06 | -0.09 |
| CGCM3.1(T47)* | -0.005 ± 0.005 | -0.002 | 0.53 | -0.36 | -0.03 |
| CNRM-CM3 | -0.752 ± 0.013 | -0.027 | 0.48 | -5.51 | -3.69 |
| CSIRO-MK3.0 | 0.007 ± 0.006 | 0.004 | 0.28 | 1.52 | 0.03 |
| ECHAM5-MPI-OM | -0.053 ± 0.007 | -0.001 | 0.40 | -0.16 | -0.26 |
| ECHO-G** | 0.053 ± 0.006 | 0.006 | 0.20 | 2.93 | 0.26 |
| FGOALS-g1.0 | -1.339 ± 0.020 | -0.044 | -0.02 | 199.19 | -6.56 |
| GFDL-CM2.0 | 0.014 ± 0.005 | -0.005 | 0.16 | -3.34 | 0.07 |
| GISS-EH | -0.013 ± 0.004 | -0.000 | -0.02 | 0.38 | -0.06 |
| GISS-ER | -0.022 ± 0.004 | 0.001 | 0.36 | 0.29 | -0.11 |
| INM-CM3.0* | -0.065 ± 0.005 | 0.007 | 0.66 | 1.07 | -0.32 |
| IPSL-CM4 | 0.198 ± 0.005 | -0.007 | 0.56 | -1.25 | 0.97 |
| MIROC3.2(medres) | -0.019 ± 0.006 | -0.001 | 0.09 | -1.16 | -0.09 |
| MRI-CGCM2.3.2*** | -0.079 ± 0.009 | 0.009 | 0.50 | 1.84 | -0.39 |
| PCM (NCAR)*** | -0.022 ± 0.005 | 0.001 | 0.40 | 0.18 | -0.11 |
| UKMO-HadCM3 | 0.003 0.005 | -0.000 | 0.25 | -0.08 | 0.02 |
| UKMO-HadGEM1 | -0.017 ± 0.005 | 0.000 | 0.09 | 0.22 | -0.09 |
| Mean | -0.136 ± 0.009 | -0.004 | 0.32 | - | -0.70 |
| Median | -0.018 ± 0.006 | -0.000 | 0.38 | - | -0.09 |

*
BCCR-BCM2.0, 2005 Bjerknes Centre for Climate Research, Norway
CCSM3, 2005 National Center for Atmospheric Research, USA
CGCM3.1(T47), 2005 Canadian Centre for Climate Modelling and Analysis, Canada
CNRM-CM3, 2004 Météo-France/Centre National de Recherches Météorologiques, France

410  CSIRO-MK3.0, 2001 Commonwealth Scientific and Industrial Research Organisation (CSIRO) Atmospheric
411  Research, Australia
412  ECHAM5/MPI-OM, 2005 Max Planck Institute for Meteorology, Germany
413  ECHO-G, 2004 Meteorological Institute of the University of Bonn, Germany
414  FGOALS-g1.0, 2004 National Key Laboratory of Numerical Modeling for Atmospheric Sciences and
415  Geophysical Fluid Dynamics (LASG)/Institute of Atmospheric Physics, China
416  GFDL-CM2.0, 2005 U.S. Department of Commerce/National Oceanic and Atmospheric Administration
417  (NOAA)/Geophysical Fluid Dynamics Laboratory (GFDL), USA
418  GISS-EH, 2004 NASA/GISS, USA
419  GISS-ER, 2004 NASA/GISS, USA
420  INM-CM3.0, 2004 Institute for Numerical Mathematics, Russia
421  IPSL-CM4, 2005 Institut Pierre Simon Laplace, France
422  MIROC3.2(medres), 2004 Center for Climate System Research (University of Tokyo), National Institute for
423  Environmental Studies, and Frontier Research Center for Global Change (JAMSTEC), Japan
424  MRI-CGCM2.3.2, 2003 Meteorological Research Institute, Japan
425  PCM, 1998 National Center for Atmospheric Research, USA
426  UKMO-HadCM3, 1997 Hadley Centre for Climate Prediction and Research/Met Office, UK
427  UKMO-HadGEM1, 2004 Hadley Centre for Climate Prediction and Research/Met Office, UK
428

Global water cycle assessment in CMIP3 climate models

**Table 2.** Global annual means, inter-annual variability and trends of atmospheric moisture transport from ocean to land areas as described in (5) for CMIP3 climate models. Listed are global annual means and standard deviations for the model time series of the combined 20$^{th}$ and 21$^{st}$ scenario A2. Further listed is the area of the dry zones as fractions of the area of the globe for the first 20-year period of the 21$^{st}$ century scenario A2. The difference in global area fraction of the dry zones between the first (2001-2020) and the last (2081-2100) 20-year period are also listed.

| CMIP3 Model | Mean ± std.dev. Atm. moisture transport ocean-land (Sv) | Trend Atm. moisture transport ocean-land (Sv/100yr) | Mean (2001 – 2020) Global dry zone area fraction | Difference (2001-2020) – (2081-2100) Global dry zone area fraction |
|---|---|---|---|---|
| BCCR-BCM2.0 | 1.01 ± 0.04 | 0.04 | 0.406 | -0.005 |
| CCSM3 | 1.44 ± 0.05 | 0.13 | 0.391 | -0.002 |
| CGCM3.1(T47)* | 1.32 ± 0.04 | 0.06 | 0.406 | -0.002 |
| CNRM-CM3 | 0.55 ± 0.05 | -0.01 | 0.385 | -0.006 |
| CSIRO-MK3.0 | 1.02 ± 0.04 | 0.02 | 0.411 | -0.005 |
| ECHAM5-MPI-OM | 0.96 ± 0.05 | 0.04 | 0.395 | -0.030 |
| ECHO-G** | 1.21 ± 0.06 | 0.11 | 0.365 | 0.009 |
| FGOALS-g1.0 | 0.26 ± 0.06 | -0.05 | No data | No data |
| GFDL-CM2.0 | 1.25 ± 0.06 | 0.01 | 0.403 | -0.012 |
| GISS-EH | 1.18 ± 0.03 | 0.02 | No data | No data |
| GISS-ER | 1.78 ± 0.05 | 0.08 | 0.398 | -0.011 |
| INM-CM3.0* | 1.11 ± 0.05 | 0.09 | 0.404 | 0.008 |
| IPSL-CM4 | 1.44 ± 0.04 | 0.08 | 0.409 | 0.007 |
| MIROC3.2(medres) | 1.20 ± 0.07 | 0.02 | 0.378 | 0.012 |
| MRI-CGCM2.3.2*** | 1.29 ± 0.12 | 0.01 | 0.401 | 0.004 |
| PCM (NCAR)*** | 1.12 ± 0.04 | 0.06 | 0.442 | 0.001 |
| UKMO-HadCM3 | 1.25 ± 0.06 | 0.08 | 0.391 | -0.004 |
| UKMO-HadGEM1 | 1.27 ± 0.06 | 0.02 | 0.417 | -0.008 |
| Mean | 1.15 ± 0.07 | 0.05 | 0.400 | -0.002 |
| Median | 1.21 ± 0.06 | 0.04 | 0.403 | -0.006 |

436

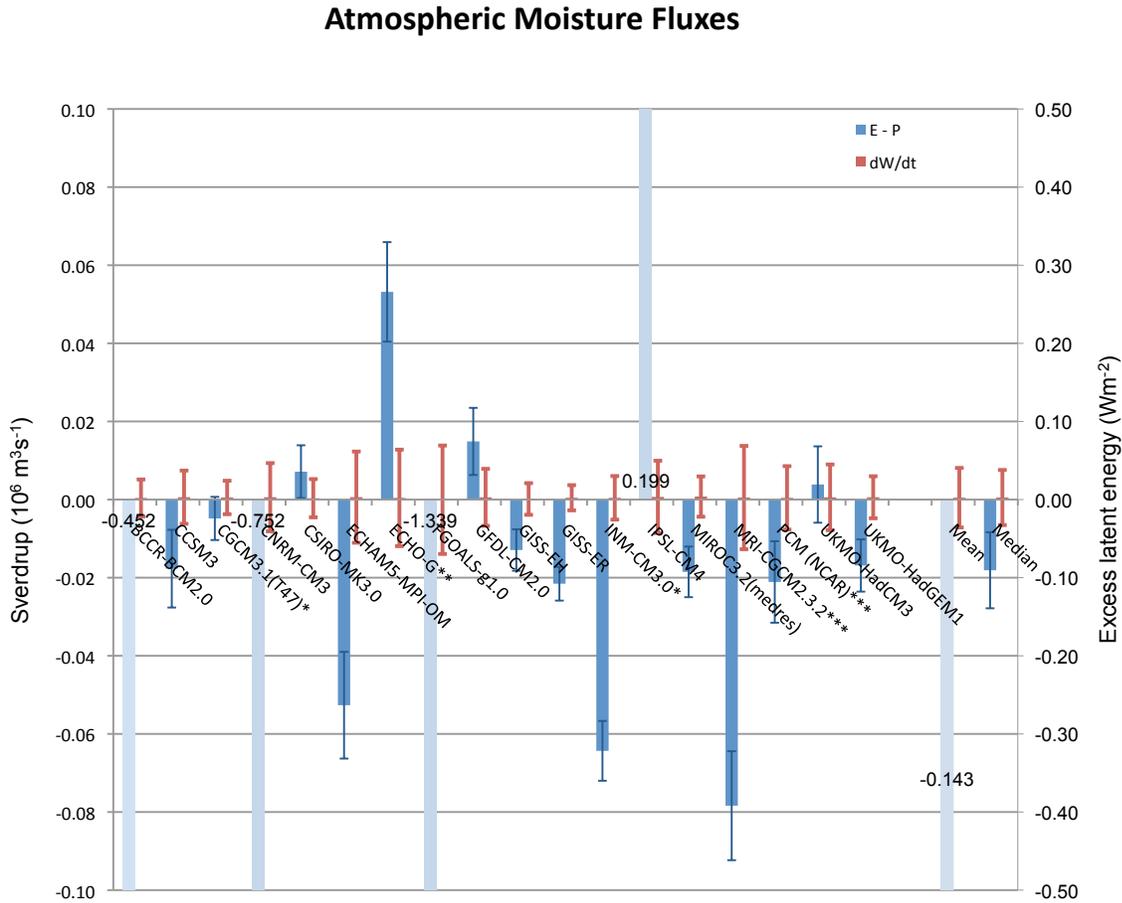

437

438 **Figure 1.** Global atmospheric moisture fluxes ***E - P*** in blue and moisture content change $\frac{\partial W}{\partial t}$ in red

439 for CMIP3 climate models. Shown are long-term annual means in columns and standard deviations in

440 error bars. The calculations were performed with the time series of the 20$^{th}$ and 21$^{st}$ scenarios A2. The

441 light blue columns are too large to be shown but the corresponding values are given on the column.

442 Note that models with flux correction are marked with *, without cloud ice with **, and without cloud

443 ice and water with ***.

444

Global water cycle assessment in CMIP3 climate models

444

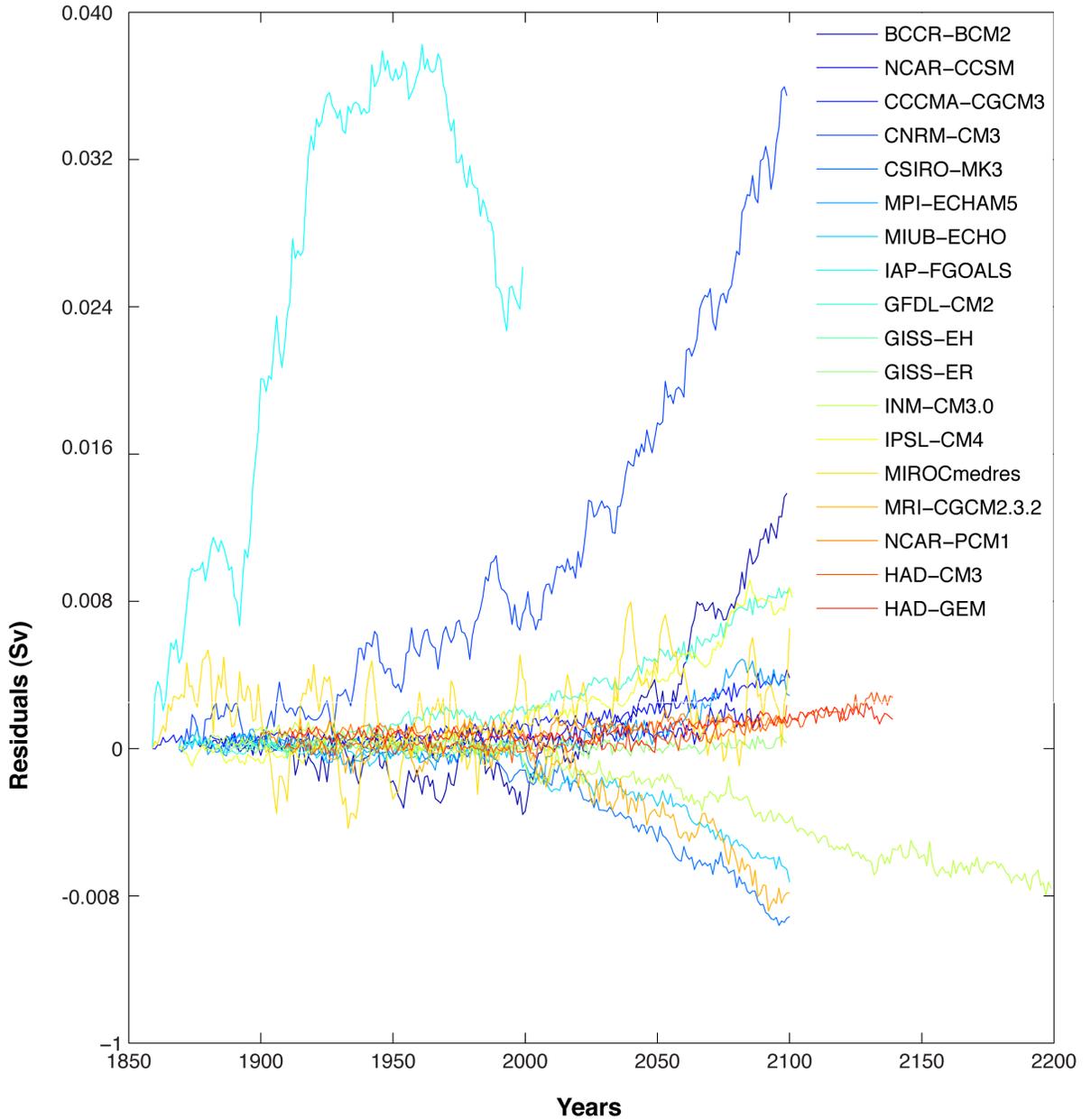

445

446 **Figure 2.** Global atmospheric moisture imbalances in CMIP3 climate models. Shown are time series of

447 annual means for the 20th and 21st scenarios A2 combined. The initial imbalances were removed from

448 the time series. The residuals are given in Sverdrup.

449

Global water cycle assessment in CMIP3 climate models

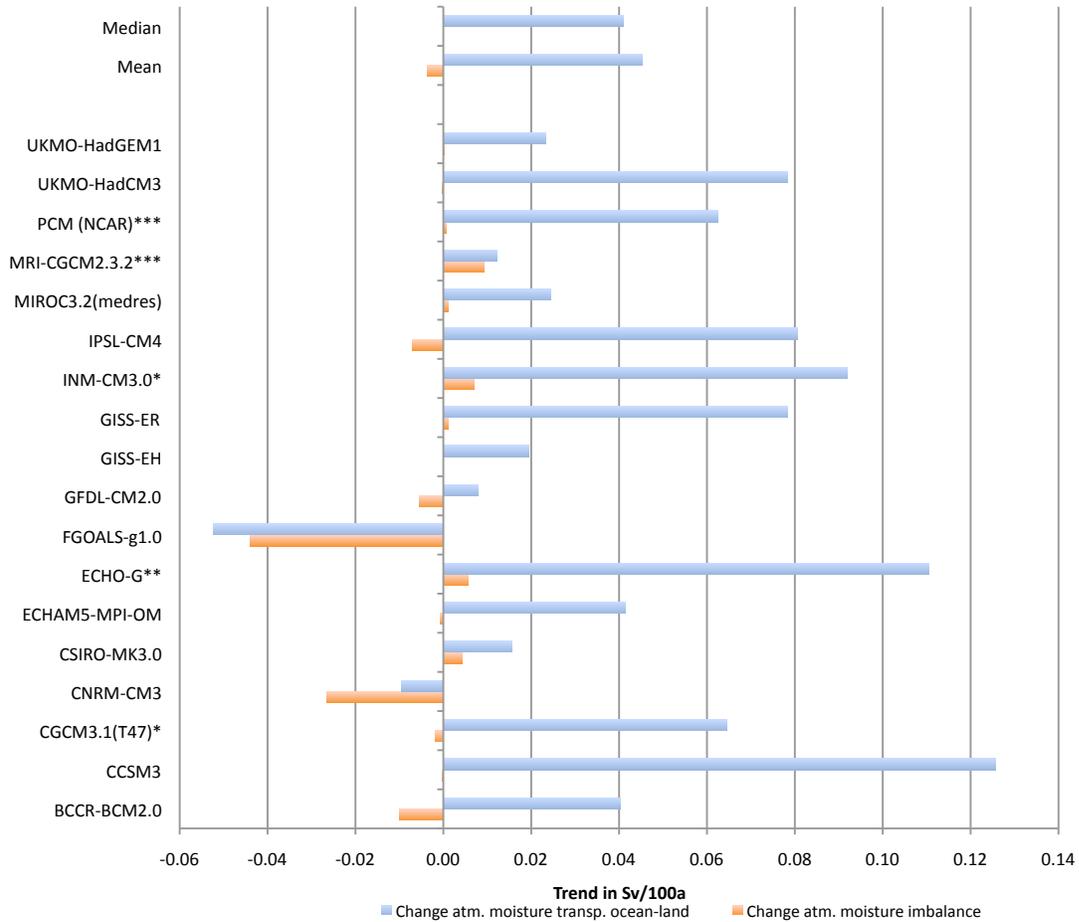

**Figure 3.** Simulated future changes in global atmospheric moisture transport from ocean to land for CMIP3 climate models. Shown are linear trends of annual mean moisture transport in blue bars and linear trends of annual mean residuals in the atmospheric moisture budgets in orange bars. The calculations were performed with the combined time series of the $20^{th}$ and $21^{st}$ (A2) century scenarios.

Global water cycle assessment in CMIP3 climate models

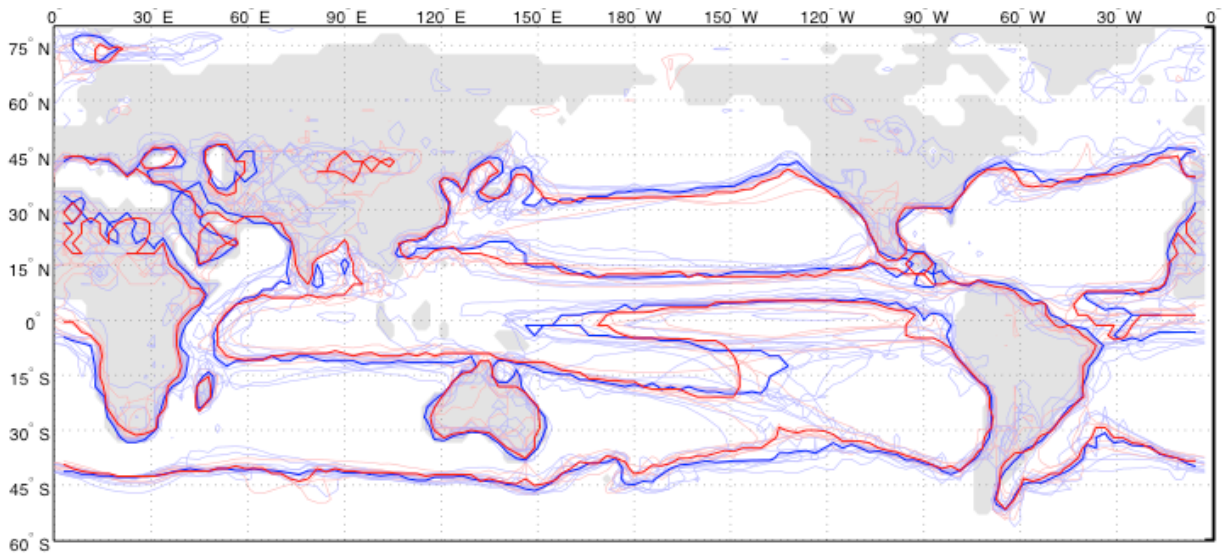

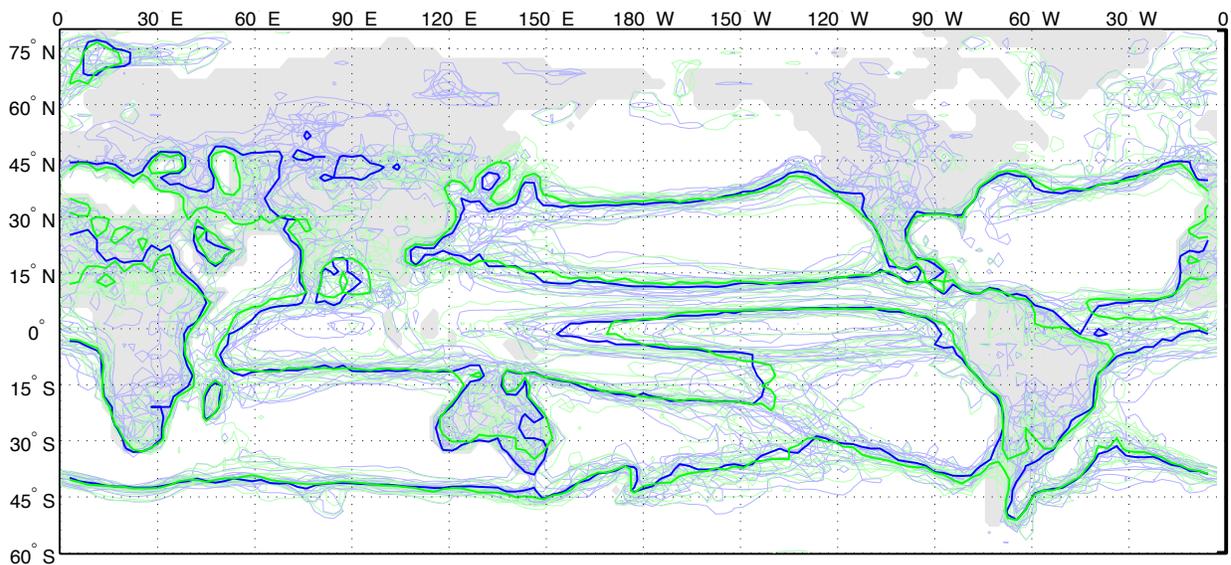

**Figure 4.** Predicted long-term mean positions of dry zone edges of the 21st century (A2) scenario for CMIP3 climate models. Top) 21st century multi-model means for two subsets of models. In red are the contours of models with artificial leaking (globally negative atmospheric moisture imbalance) and in blue with artificial flooding of the atmospheres (globally positive atmospheric moisture imbalance). Bottom) 21st century multi-model means for two 20-year periods. In green are the contours of the dry-zone edges from the means of the last two (2081 – 2100) and in blue for the first two decades (2001 – 2020). Shown are the contour lines of balanced atmospheric moisture budget, which corresponds to $E = P$, where evaporation equals precipitation. The thick contour lines represent the multi-model composites of dry-zone edges of the corresponding subsets.

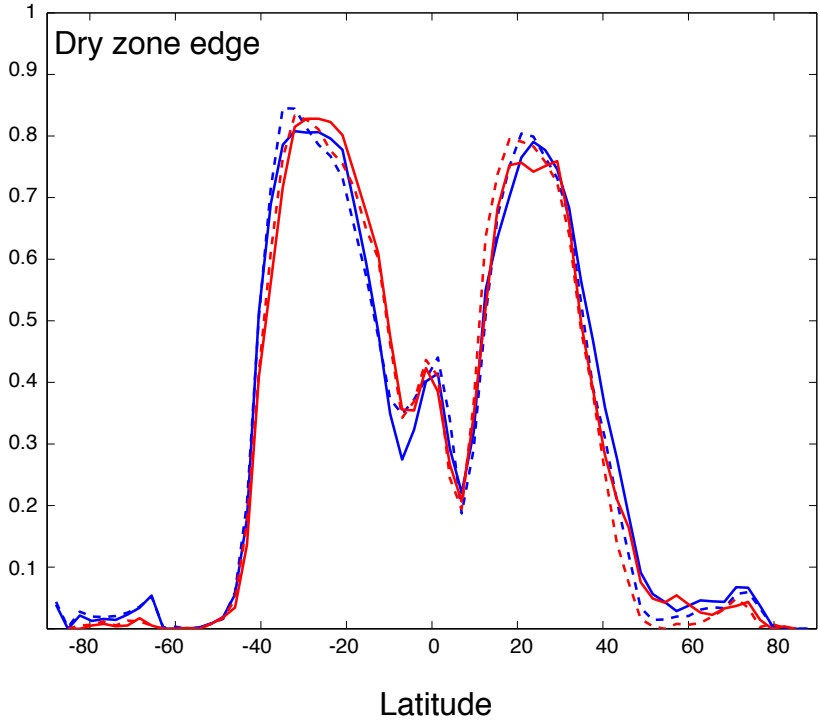

**Figure 5.** The four curves represent the zonal mean positions of CMIP3 multi-model mean dry zone edges of the first (full line) and last (dashed line) 20-year periods of the 21$^{st}$ century A2 scenario. The multi-model means are separated into subsets of artificially leaking (globally negative atmospheric moisture imbalance) models in red and flooding (globally positive atmospheric moisture imbalance) models in blue. The zonal means are calculated for the atmospheric moisture budgets, with value 'zero' set for grid boxes where $E < P$, and value 'one' set $E \geq P$.